\documentclass[11pt]{article}
\pdfoutput=1  
\usepackage{graphicx,color}
\usepackage{appendix}
\usepackage{latexsym,amsmath,amssymb,graphicx,booktabs}
\usepackage{epsfig,latexsym,cite}
\usepackage{hyperref}
\numberwithin{equation}{section}

\definecolor{MyBlue}{rgb}{0.15,0.15,0.70}

\hypersetup{
colorlinks=true,
citecolor=MyBlue,
linkcolor=MyBlue,
urlcolor=MyBlue
}

\input epsf
\usepackage{amssymb}
\usepackage{amsmath}
\usepackage{amsfonts}
\usepackage{upgreek}
\usepackage{latexsym}

\newcommand{\iBox}{\Box^{-1}}

\renewcommand\({\left(}
\renewcommand\){\right)}

\renewcommand\]{\right]}

\def\lsim{\raise 0.4ex\hbox{$<$}\kern -0.8em\lower 0.62
ex\hbox{$\sim$}}

\def\gsim{\raise 0.4ex\hbox{$>$}\kern -0.7em\lower 0.62
ex\hbox{$\sim$}}

\def\lbar{{\hbox{$\lambda$}\kern -0.7em\raise 0.6ex
\hbox{$-$}}}

\newcommand\eq[1]{Eq.~(\ref{#1})}

\newcommand\p{\partial}

\newcommand\ee{\end{equation}}
\newcommand\be{\begin{equation}}
\def\bea{\begin{array}}
\def\eea{\end{array}}\def\ea{\end{array}}
\newcommand\ees{\end{eqnarray}}
\newcommand\bees{\begin{eqnarray}}

\def\s{\sigma}

\def\dslash{\hspace{-1mm}\not{\hbox{\kern-2pt $\partial$}}}
\def\Dslash{\not{\hbox{\kern-4pt $D$}}}
\def\pslash{\not{\hbox{\kern-2.1pt $p$}}}
\def\kslash{\not{\hbox{\kern-2.3pt $k$}}}
\def\qslash{\not{\hbox{\kern-2.3pt $q$}}}

\def\p1{{\bf p}_1}
\def\p2{{\bf p}_2}
\def\k1{{\bf k}_1}
\def\k2{{\bf k}_2}

\newcommand{\gmn}{g_{\mu\nu}}

\newcommand{\Gmn}{G_{\mu\nu}}

\newcommand{\Tmn}{T_{\mu\nu}}

\newcommand{\dddM}{\kern 0.2em \raise 1.9ex\hbox{$...$}\kern -1.0em \hbox{$M$}}
\newcommand{\dddQ}{\kern 0.2em \raise 1.9ex\hbox{$...$}\kern -1.0em \hbox{$Q$}}
\newcommand{\dddI}{\kern 0.2em \raise 1.9ex\hbox{$...$}\kern -1.0em\hbox{$I$}}
\newcommand{\dddJ}{\kern 0.2em \raise 1.9ex\hbox{$...$}\kern-1.0em
\hbox{$J$}}
\newcommand{\dddcalJ}{\kern 0.2em \raise 1.9ex\hbox{$...$}\kern-1.0em
\hbox{${\cal J}$}}

\newcommand{\dddO}{\kern 0.2em \raise 1.9ex\hbox{$...$}\kern -1.0em
\hbox{${\cal O}$}}
\def\dddz{\raise 1.5ex\hbox{$...$}\kern -0.8em \hbox{$z$}}
\def\dddd{\raise 1.8ex\hbox{$...$}\kern -0.8em \hbox{$d$}}
\def\dddbd{\raise 1.8ex\hbox{$...$}\kern -0.8em \hbox{${\bf d}$}}
\def\ddbd{\raise 1.8ex\hbox{$..$}\kern -0.8em \hbox{${\bf d}$}}
\def\dddx{\raise 1.6ex\hbox{$...$}\kern -0.8em \hbox{$x$}}

\newcommand{\ode}{\Omega_{\rm DE}}

\newcommand{\ola}{\Omega_{\Lambda}}

\begin{document}

\begin{titlepage}

\vspace*{2cm}

\centerline{\Large \bf Non-local gravity and comparison with observational datasets\footnote{Based on observations obtained with Planck (http://www.esa.int/Planck), an ESA science mission with instruments and contributions directly funded by ESA Member States, NASA, and Canada.}
}

%

\vskip 0.4cm
\vskip 0.7cm
\centerline{\large Yves Dirian$^a$, Stefano Foffa$^a$, Martin Kunz$^{a,b}$, Michele Maggiore$^a$ and Valeria Pettorino$^c$}
\vspace{3mm}
\centerline{\em $^a$D\'epartement de Physique Th\'eorique and Center for Astroparticle Physics,}  
\centerline{\em Universit\'e de Gen\`eve, 24 quai Ansermet, CH--1211 Gen\`eve 4, Switzerland}
\vspace{3mm}
\centerline{\em $^b$African Institute for Mathematical Sciences, 6 Melrose Road, Muizenberg, 7945, South Africa}
\vspace{3mm}
\centerline{\em $^c$Institut f\"{u}r Theoretische Physik, Universit\"{a}t Heidelberg,
Philosophenweg 16, D-69120 Heidelberg}

\vskip 1.9cm

\begin{abstract}

We study the cosmological predictions of   two recently proposed non-local modifications of General Relativity.  Both models have the same number of parameters as $\Lambda$CDM, with a mass parameter $m$ replacing the cosmological constant. We implement the cosmological perturbations of the non-local models into a modification of the CLASS  Boltzmann code, and we make a full comparison  to CMB, BAO and supernova  data. We find that the non-local models fit these datasets as well as $\Lambda$CDM.  For both non-local models parameter estimation using {\em Planck}\,+JLA+BAO data gives a value of $H_0$ higher than in $\Lambda$CDM, and in better agreement with the values obtained from local measurements.

\end{abstract}

\end{titlepage}

\newpage

\section{Introduction}

The observational evidence for the accelerated expansion of the Universe \cite{Riess:1998cb,Perlmutter:1998np} has stimulated renewed interest in modifications of General Relativity (GR). 
A possible approach, which has been suggested by different lines of investigations, is to add some non-local terms  to GR. Non-locality in this case should not be considered as fundamental. In many physical situations non-local terms emerge from a fundamental local theory, by a classical or a quantum averaging process. For instance, non-local (but causal) effective equations govern  the  dynamics of the in-in matrix elements of quantum fields,  and encode  ultraviolet (UV) quantum corrections  
to the classical dynamics \cite{Jordan:1986ug,Calzetta:1986ey}. The cosmological consequences of  non-local UV effects have been recently  studied e.g. in \cite{Biswas:2010zk,Biswas:2011ar,Biswas:2013kla,Donoghue:2014yha}. UV effects are however expected to be relevant only in the large-curvature regime, so for instance for the issue of smoothing the big-bang singularity, but  should not be cosmologically relevant in the present epoch.
Non-local modifications of GR  are however also  expected to emerge from infrared (IR) corrections to the effective field equations. These  are indeed known to become potentially large in quantum field theory in curved space, most notably in de~Sitter, which is the most studied case, see e.g. \cite{Mottola:1984ar,Tsamis:1996qm,Tsamis:1996qq,Sahni:1998at,Polyakov:2007mm,Burgess:2009bs,Burgess:2010dd,Polyakov:2012uc,Dolgov:2005se,Akhmedov:2013vka,Tsamis:2011ep,Anderson:2013ila}, and therefore they can potentially modify the long-distance behavior of GR.

Ultraviolet  corrections in quantum field theory in curved space are by now well understood, as summarized in textbooks such as \cite{Birrell:1982ix}. The situation for IR effects in curved space is much more complicated. Often, they manifest themselves through secularly growing terms in back-reaction computations. Such terms signal the onset of an instability, but it is typically beyond the present technology to follow the fate of the instability when the back-reaction becomes large, and to compute from first principles a corresponding effective (and in general non-local)  equation of motion that describes these effects. While a better understanding of infrared effects in curved space would be highly desirable, a simpler  phenomenological attitude  is to postulate a non-local modification of GR which involves inverse powers of the d'Alembertian, and therefore becomes relevant in the IR, and to study its cosmological consequences.  Eventually such a program will only be successful  if one will be able to derive such non-local terms from first principles. However, a first step can be to understand what sort of non-local terms can give rise to an interesting  cosmology. Identifying a non-local model that works well with respect to the cosmological observations would  be of great help in understanding  how to derive such an effective theory from fundamental principles (much as understanding the structure of the Fermi theory of weak interactions at low energies was instrumental for building the Standard Model, several decades afterwards).
 
In this spirit, in recent years there have been many investigations of non-local modifications of GR. For instance, non-local operators appear in the degravitation proposal \cite{ArkaniHamed:2002fu,Dvali:2006su}, where the insertion in the Einstein equations of an operator of the form $(1-m^2/\Box)$ was argued to have a screening effect on the cosmological constant (see also
\cite{Patil:2010mq}). Non-local long-distance modifications of GR have been suggested in \cite{Barvinsky:2003kg,Barvinsky:2011hd,Barvinsky:2011rk,Wetterich:1997bz}.
Constructing a non-local model that produces a dynamical dark energy and fits well the observations is however quite non-trivial. For instance, in recent years much attention has been devoted to a 
non-local cosmological model   proposed  in 
\cite{Deser:2007jk} (see \cite{Koivisto:2008xfa,Koivisto:2008dh,Capozziello:2008gu,Elizalde:2011su,Zhang:2011uv,Elizalde:2012ja,Park:2012cp,Bamba:2012ky,Deser:2013uya,Ferreira:2013tqn,Dodelson:2013sma,Woodard:2014iga}).  
This model is based on the addition of a term of the form $Rf(\iBox R)$ to the Einstein-Hilbert action, where $R$ is the Ricci scalar. The function $f(X)$ was chosen  to obtain  a viable cosmology for the background evolution. The result turns out to be not very natural,  $f(X)=a_1[\tanh (a_2Y+a_3Y^2+a_4Y^3)-1]$, where $Y=X+a_5$, and $a_1,\ldots a_5$ are coefficients fitted to the observed expansion history. More importantly, once the function $f(X)$ is fixed in this way, one can compute the cosmological perturbations  and it turns out that this model is ruled out with great statistical significance,   at  7.8$\sigma$  from redshift space distortions, and at  5.9$\s$  from weak lensing~\cite{Dodelson:2013sma}.
A different non-local approach, which  appears to be phenomenologically successful,  has been recently developed by our group~\cite{Jaccard:2013gla,Maggiore:2013mea,Foffa:2013sma,Foffa:2013vma,Kehagias:2014sda,Maggiore:2014sia,Dirian:2014ara,Cusin:2014zoa,Dirian:2014xoa} and further discussed in
\cite{Modesto:2013jea,Nesseris:2014mea,Conroy:2014eja,Barreira:2014kra}. In its simplest form, it is based on the action\cite{Maggiore:2014sia}
\be\label{S1}
S_{\rm NL}=\frac{1}{16\pi G}\int d^{4}x \sqrt{-g}\, 
\[R-\frac{1}{6} m^2R\frac{1}{\Box^2} R\]\, .
\ee
Integrating by parts  the  $\iBox$ operator,
this non-local action can be rewritten as
\be\label{S1fX}
S_{\rm NL}=\frac{1}{16\pi G}\int d^{4}x \sqrt{-g}\, 
\[R-\frac{1}{6} m^2f(\iBox R)\]\, ,
\ee
where $f(X)= X^2$. This model  works well, compared to  cosmological observations, both at the level of background evolution and at the level of cosmological perturbations. 
The fact that this success is obtained with the simple choice $f(X)= X^2$, rather than with a highly fine-tuned function, certainly makes the model stand out for its simplicity. Nonetheless, it is  important to explore also different related non-local models, to see to what extent one can extract  general predictions.  From this point of view, a first useful observation is that models involving tensor non-localities, e.g. involving terms such as $R_{\mu \nu} \Box^{-2} R^{\mu \nu}$ in the action, or terms $\iBox R_{\mu \nu}$ in the equations of motion, do not provide a viable cosmological evolution already at the background level, since they are plagued by fatal run-away instabilities~\cite{Maggiore:2014sia,Modesto:2013jea,Foffa:2013vma,Ferreira:2013tqn}. This restricts significantly the class of viable models, and provides potentially useful indications for the construction of the corresponding fundamental theory.

To understand how much the results depend on our choice of a specific model, in our previous studies we have compared the model (\ref{S1}) to another non-local model, which is  defined directly at the level of equations of motion, by \cite{Maggiore:2013mea} 
\be\label{modelRT}
\Gmn -\frac{1}{3}m^2\(\gmn\iBox R\)^{\rm T}=8\pi G\,\Tmn\, ,
\ee
where the inverse of the d'Alembertian is defined with the retarded Green's function, and
the superscript ``T" denotes the extraction of the transverse part of a tensor (which is itself a non-local operation). The extraction of the transverse part  ensures that the left-hand side
of \eq{modelRT} has zero divergence, and therefore $\Tmn$ is automatically conserved. 
While the model defined by \eq{S1} corresponds to the simplest possible action in this class, the model (\ref{modelRT}) provides the  most compact equation of motion, so in a sense they are both selected by simplicity. These two models are also related by the fact that, when linearizing the equations of motion derived from the action (\ref{S1}) around flat space, one finds the same equations of motion as those obtained by linearizing \eq{modelRT} \cite{Maggiore:2014sia}. However, beyond the linear level in an expansion over Minkowski space, or for generic  backgrounds 
(such as FRW), the equations of motion of the two theories are different.

In the end, if the comparison with the data should eventually point toward the correctness of a non-local model of this sort, it might also point toward the necessity of  refining it. It must certainly be borne in mind that our quantitative results are in any case specific to the models that we use.
Ideally, one would like to eventually derive the non-local model from first principles, and  this should  select the exact non-local structure. The purely phenomenological approach that we rather take here could   help in identifying a promising non-local structure,
paving the way for a first-principle approach.

\begin{table}[b]
\begin{center}
\begin{tabular}{|l|c|c|c|} 
 \hline 
 Param & $\Lambda$CDM  & $g_{\mu\nu}\Box^{-1}R$ &$R\Box^{-2}R$\\ \hline 
$100~\omega_{b }$  & $2.201_{-0.029}^{+0.028}$&$2.204_{-0.03}^{+0.028}$
& $2.207_{-0.029}^{+0.029}$ \\ 
$\omega_c$  & $0.1194_{-0.0026}^{+0.0027}$& $0.1195_{-0.0028}^{+0.0026}$
& $0.1191_{-0.0028}^{+0.0027}$\\ 
$H_0$  & $67.56_{-1.3}^{+1.2}$& $68.95_{-1.3}^{+1.3}$& $71.67_{-1.5}^{+1.5}$\\ 
$10^{9}A_{s}$ & $2.193_{-0.06}^{+0.052}$& $2.194_{-0.062}^{+0.048}$
& $2.198_{-0.059}^{+0.053}$\\ 
$n_{s}$  & $0.9625_{-0.0074}^{+0.0072}$& $0.9622_{-0.0081}^{+0.007}$
& $0.9628_{-0.0073}^{+0.0074}$\\ 
$z_{re}$  & $11.1_{-1.1}^{+1.1}$& $11.1_{-1.2}^{+1.1}$ 
& $11.16_{-1.1}^{+1.2}$\\  
\hline
$\chi^2_{\rm min}$ &$9801.7$ &$9801.3$&9800.1\\
\hline
\end{tabular}  
 \end{center}
\caption{$68\%$ limits for the cosmological parameters of $\Lambda$CDM and of the two non-local models, using the {\em Planck\,} CMB data only. The last row gives the  minimum chi-square of the best fit. $H_0$ is in units of ${\rm km}\,{\rm s}^{-1}{\rm Mpc}^{-1}$.\label{tab1}}
\end{table}

In our previous studies of the models (\ref{S1}) and (\ref{modelRT}) we found that they give qualitatively similar cosmological predictions~\cite{Maggiore:2013mea,Maggiore:2014sia,Dirian:2014ara}. In particular, in both models the non-local term  effectively acts as  a dynamical dark energy (DE), that can explain the present acceleration of the Universe. Furthermore,  for very general reasons they both predict a phantom DE equation of state. Numerical details of course differ. After fixing the mass $m$ so as to reproduce the observed value of $\ode$, the model (\ref{S1}) predicts that, today,
$w_{\rm DE}\simeq -1.14$, while \eq{modelRT} 
predicts $w_{\rm DE}\simeq -1.04$. 
 Since cosmological perturbations in the DE sector are mostly proportional to $(1+w_{\rm DE})$, we also generically find that the predictions of 
the model (\ref{modelRT}) are intermediate between that of model (\ref{S1}) and that of 
$\Lambda$CDM.

The aim of this paper is to perform a detailed comparison of these non-local models  with  cosmological observations. 
We will refer to them as the ``$R\,\Box^{-2}R$ model" and the ``$\gmn\iBox R$ model", respectively.
For the  $\gmn\iBox R$ model,  cosmological perturbations 
have already been worked out in \cite{Nesseris:2014mea,Dirian:2014ara}. In particular, Nesseris and Tsujikawa  \cite{Nesseris:2014mea} showed  that the  $\gmn\iBox R$ model is consistent witn SNe (Union 2.1), BAO, CMB and growth rate data. However for the CMB they only used the Planck CMB shift parameters and did not implement the model in a Boltzmann code. For the $R\,\Box^{-2}R$ model, cosmological perturbations have been worked out in 
\cite{Dirian:2014ara}, where again consistency with SN and structure formation data was found.
However,  an accurate comparison with CMB data requires to implement  the cosmological perturbations of these models in a Boltzmann code. 
We have now implemented the cosmological perturbations of both non-local models in a Boltzmann code, modifying the CLASS code \cite{Lesgourgues:2011re,Blas:2011rf,Lesgourgues:2011rg,Lesgourgues:2011rh}. In this paper we present an accurate  comparison of these models with CMB, SNe, BAO and HST data. We  perform parameter estimation for these models, and we compare their goodness of fit to that of $\Lambda$CDM, using the Markov Chain Monte Carlo (MCMC) code Montepython v2.1.0~\cite{Audren:2012wb}.
In this paper we present the main results of this analysis. 
A more extended  discussion will be presented elsewhere.

\begin{table}[t]
\begin{center}
\begin{tabular}{|l|c|c|c|} 
 \hline 
Param & $\Lambda$CDM  & $g_{\mu\nu}\Box^{-1}R$ &$R\Box^{-2}R$\\ \hline 
$100~\omega_{b }$  & $2.215_{-0.025}^{+0.025}$
&$2.207_{-0.025}^{+0.024}$
& $2.197_{-0.025}^{+0.024}$ \\ 
$\omega_c$  & $0.1175_{-0.0014}^{+0.0015}$
& $0.1188_{-0.0014}^{+0.0014}$
& $0.1204_{-0.0013}^{+0.0014}$\\ 
$H_0$  & $68.43_{-0.69}^{+0.61}$
& $69.3_{-0.66}^{+0.68}$
& $70.94_{-0.7}^{+0.74}$\\ 
$10^{9}A_{s}$ & $2.199_{-0.062}^{+0.055}$
& $2.196_{-0.065}^{+0.052}$
& $2.192_{-0.061}^{+0.051}$\\ 
$n_{s}$  & $0.9668_{-0.0054}^{+0.0055}$
& $0.9636_{-0.0055}^{+0.0052}$
& $0.9599_{-0.0051}^{+0.0052}$\\ 
$z_{re}$  & $11.33_{-1.1}^{+1.1}$
& $11.18_{-1.2}^{+1.1}$
& $11.00_{-1.2}^{+1.1}$\\ 
\hline 
$\chi^2_{\rm min}$ &$10485.5$ 
&$10485.0$
&10488.7\\
\hline
\end{tabular}  
\end{center}
\caption{As Table~\ref{tab1}, using {\em Planck\,}+JLA+BAO data.
\label{tab2}}
\end{table}

\begin{table}[t]
\begin{center}
\begin{tabular}{|l|c|c|c|} 
 \hline 
Param & $\Lambda$CDM  & $g_{\mu\nu}\Box^{-1}R$ &$R\Box^{-2}R$\\ \hline 
$100~\omega_{b }$  & $2.222_{-0.025}^{+0.025}$
&$2.212_{-0.025}^{+0.024}$
& $2.202_{-0.024}^{+0.023}$ \\ 
$\omega_c$  & $0.117_{-0.0014}^{+0.0014}$
& $0.1182_{-0.0014}^{+0.0013}$
& $0.1201_{-0.0013}^{+0.0013}$\\ 
$H_0$  & $68.72_{-0.63}^{+0.61}$
& $69.60_{-0.63}^{+0.66}$
& $71.14_{-0.69}^{+0.72}$\\ 
$10^{9}A_{s}$ & $2.202_{-0.067}^{+0.053}$
& $2.198_{-0.059}^{+0.053}$
& $2.195_{-0.058}^{+0.053}$\\ 
$n_{s}$  & $0.9679_{-0.0054}^{+0.0052}$
& $0.9649_{-0.0056}^{+0.0052}$
& $0.9607_{-0.0050}^{+0.0051}$\\ 
$z_{re}$  & $11.39_{-1.3}^{+1.1}$
& $11.25_{-1.1}^{+1.1}$
& $11.05_{-1.1}^{+1.1}$\\ 
\hline 
$\chi^2_{\rm min}$ &$10488.9$
 &$10487.3$
&10489.3\\
\hline
\end{tabular} 
\end{center} 
\caption{As Table~\ref{tab1}, adding to {\em Planck\,}+JLA+BAO also the HST value $H_0=73.0\pm 2.4$.
\label{tab3}}
\end{table}

\begin{figure}[t]
\centering
\includegraphics[width=0.7\columnwidth]{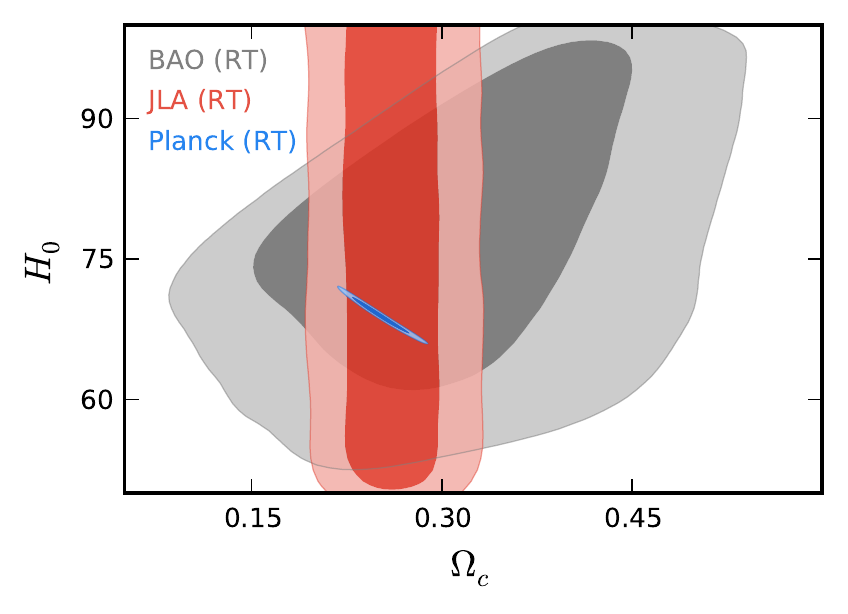}
\includegraphics[width=0.7\columnwidth]{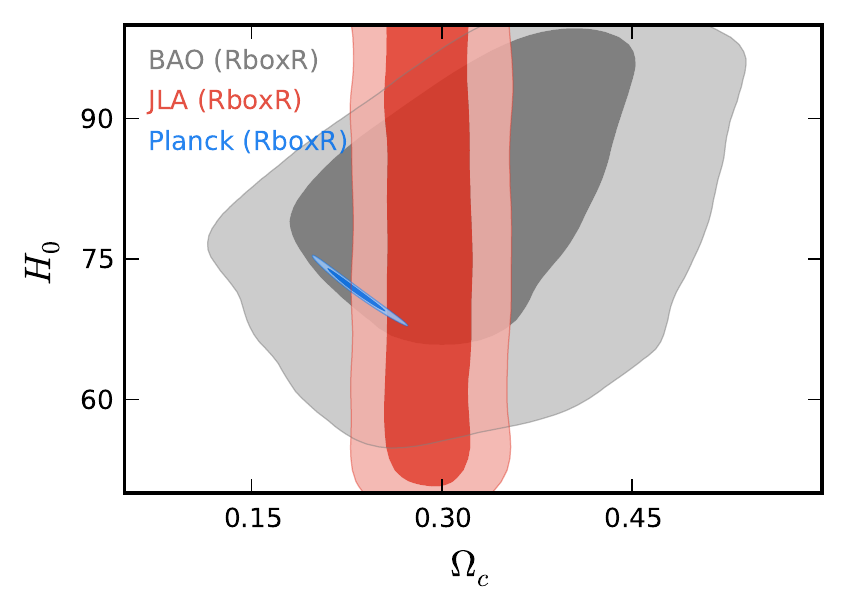}
\caption{\label{contours} The separate $1\sigma$ and $2\sigma$ contours for CMB, BAO and SNe in the plane
$(H_0,\Omega_c)$. Top:   the $g_{\mu\nu}\Box^{-1}R$ model (labeled RT). Bottom: the $R\Box^{-2}R$ model
(labeled RboxR).
}
\end{figure}

\section{Results} We use as datasets the 
CMB data from the {\em Planck\,} 2013 data release \cite{Ade:2013zuv},  type-Ia supernovae  from 
JLA \cite{Betoule:2014frx}, and BAO data from BOSS
\cite{Anderson:2013zyy} and 6dF \cite{Beutler:2011hx}.
As a set of cosmological parameters we vary in our analysis
 the baryon density today $\omega_b=\Omega_bh_0^2$,  the cold dark matter density
$\omega_c=\Omega_{\rm c}h_0^2$, the Hubble parameter today 
$H_0=h_0(100\,{\rm km}\,{\rm s}^{-1}{\rm Mpc}^{-1})
$, the amplitude of scalar perturbation $A_s$, the scalar spectrum index $n_s$ and the redshift at which the Universe is half-reionized $z_{\rm re}$. 
In $\Lambda$CDM the dark energy density $\ola$ is a derived parameter, fixed by the flatness condition. Similarly, in our model the mass parameter $m^2$ is a derived parameter, fixed again from the condition  $\Omega_{\rm tot}=1$. The non-local models and $\Lambda$CDM therefore have the same set of free parameters, which facilitates their comparison.

\begin{figure}[t]
\centering
\includegraphics[width=0.8\columnwidth]{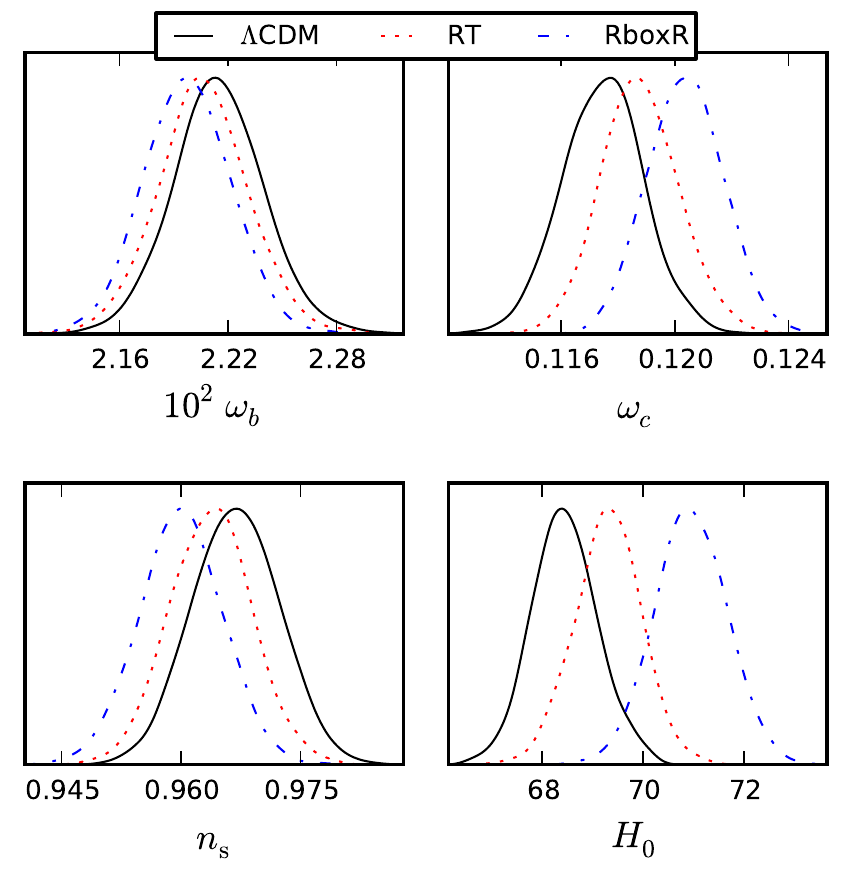}
\caption{\label{fig:1D} The likelihood for  $\omega_b$,  $\omega_c$, $n_s$ and $H_0$, for the {\em Planck}+BAO+JLA datasets, for $\Lambda$CDM (solid black line), $g_{\mu\nu}\Box^{-1}R$ (dotted red) and
$R\Box^{-2}R$ (dot-dashed, blue). 
}
\end{figure}

\begin{figure}[h]
\centering
\includegraphics[width=0.8\columnwidth]{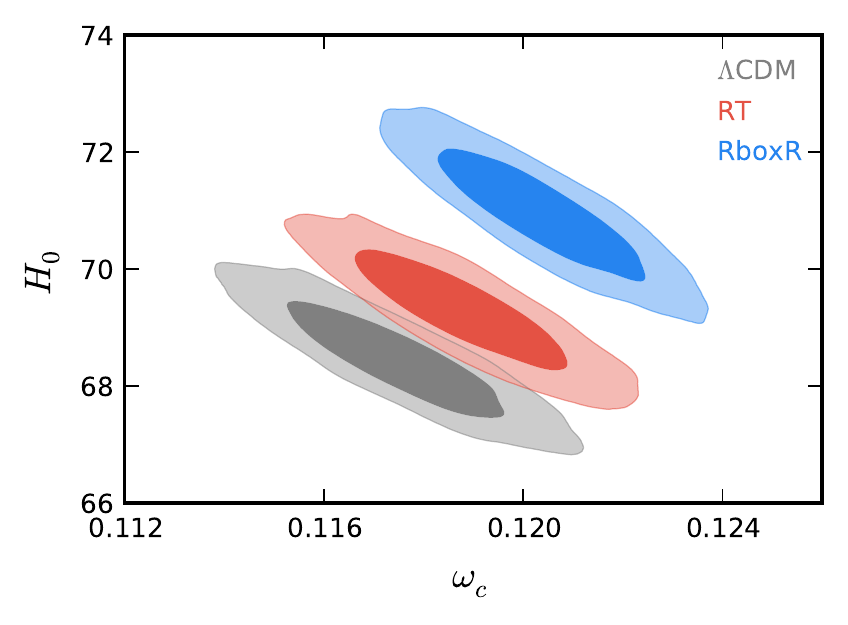}
\caption{\label{fig:H0Omega} The $1\sigma$ and $2\sigma$ contours  in the
$(H_0,\omega_c)$ plane for the three models, using {\em Planck\,}+JLA+BAO.}
\end{figure}

Table~\ref{tab1} shows the mean values with the $1\s$ errors for these parameters, obtained from our  MCMC using the {\em Planck\,} data only. In Table~\ref{tab2} we show the results obtained combining {\em Planck,}  JLA  and BAO. Our values for $\Lambda$CDM are in agreement, within the statistical error, with those reported in
\cite{Ade:2013zuv}. We see from the tables that 
the $g_{\mu\nu}\Box^{-1}R$  case gives intermediate predictions  between $\Lambda$CDM and
the $R\Box^{-2}R$ model. Fig.~\ref{contours} shows the separate $1\sigma$ and $2\sigma$ contours for CMB, BAO and SNe in the plane
$(H_0,\Omega_c)$. We notice  that for the $g_{\mu\nu}\Box^{-1}R$ model  these dataset are fully consistent, while in the $R\Box^{-2}R$ model there is a slight tension between CMB and SN data, which explains the higher $\chi^2$ for this model in Table~\ref{tab2} (although even for this model the datasets are in agreement within $2\sigma$).

\begin{figure}[t]
\centering
\includegraphics[width=0.8\columnwidth]{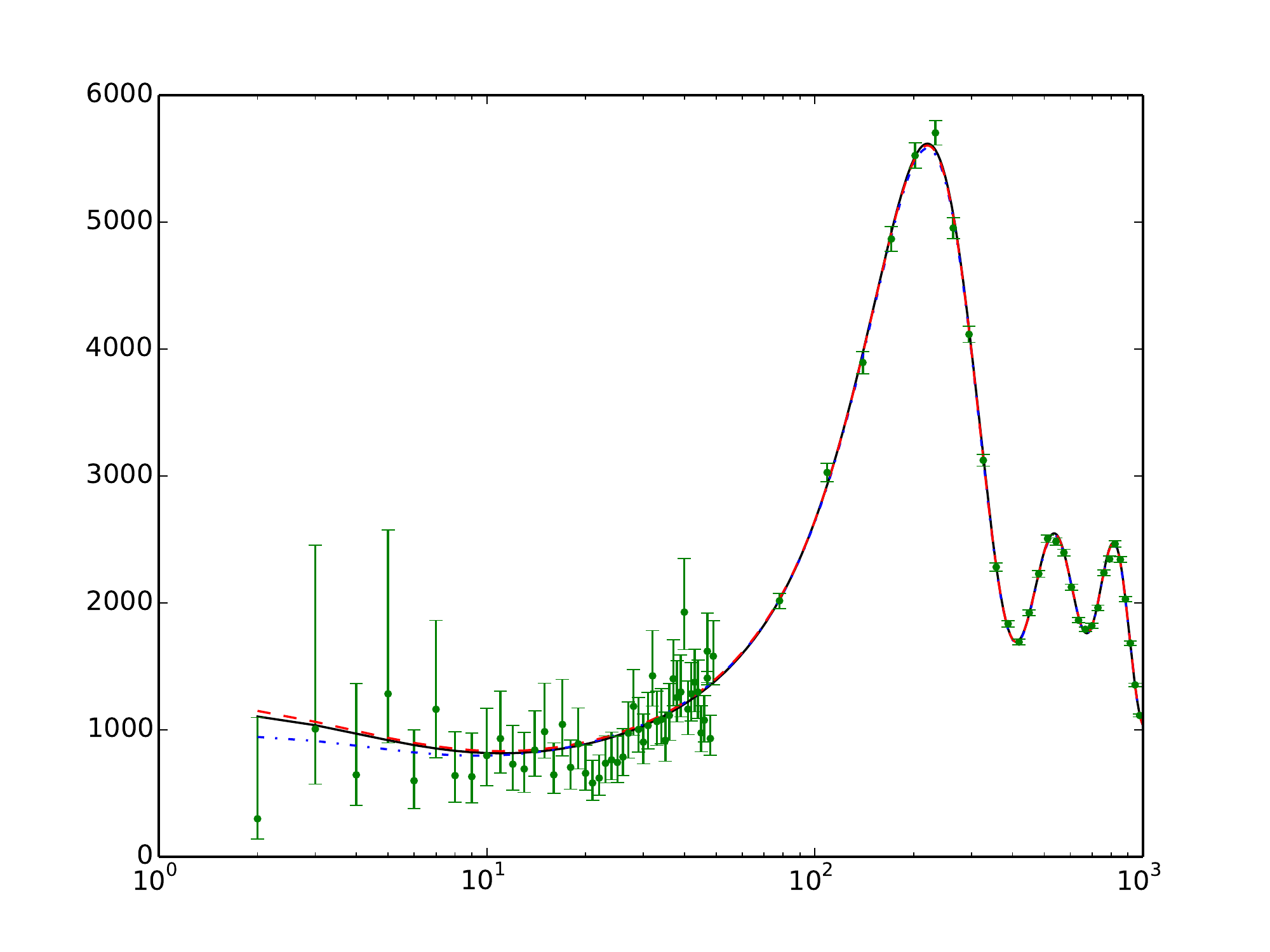}
\caption{\label{fig:Cl} A plot of the Planck data for $l(l+1)C_l/(2\pi)$, together with the  
curves  obtained  with the best-fit parameters determined fitting to the {\em Planck\,}+JLA+BAO dataset,
for $\Lambda$CDM (solid black line), the $g_{\mu\nu}\Box^{-1}R$ model (red dashed line, almost indistinguishable from the black line) and  the $R\Box^{-2}R$ model (blue, dot-dashed).
}
\end{figure}

In Fig.~\ref{fig:1D} we show the marginalized likelihood for  $\omega_b$,  $\omega_c$, $n_s$ and $H_0$, while in  Fig.~\ref{fig:H0Omega} we show the $1\sigma$ and $2\sigma$ contours of the  likelihood function in the 
$(H_0,\omega_c)$ plane,  for $\Lambda$CDM and for the two non-local models.
Among the various parameters, the most significant difference is in $H_0$, which in both non-local models is higher than in $\Lambda$CDM. Local measurements of $H_0$ from the HST give $H_0=73.8\pm 2.4$ 
(in units of ${\rm km}\,{\rm s}^{-1}{\rm Mpc}^{-1}$)
\cite{Riess:2011yx}, or $H_0=73.0\pm 2.4$ after the recalibration to NGC 4258 in \cite{Humphreys:2013eja},
while \cite{Freedman:2012ny} gives $H_0=74.3\pm 1.5 ({\rm stat})\pm 2.1 ({\rm sys})$. 
At present there is no consensus on whether these high values are due to unaccounted systematics in the SN data  \cite{Efstathiou:2013via}, statistical fluctuations \cite{Bennett:2014tka}, or  an indication of deviations from  $\Lambda$CDM. In any case, it is  interesting to observe that in the non-local models, using only the   {\em Planck\,}+JLA+BAO dataset (so, {\em without} including HST), $H_0$ automatically comes out  higher. Thus,  including also  the HST value in the fit tends to favor the non-local models over $\Lambda$CDM. In Table~\ref{tab3} we show the results of adding to the {\em Planck\,}+JLA+BAO dataset also the HST value $H_0=73.0\pm 2.4$, chosen just as an example of the impact 
of a high value. We see that the $\chi^2$ for the three model are quite comparable, with a slight preference for the
$g_{\mu\nu}\Box^{-1}R$ model.

Finally, in Fig.~\ref{fig:Cl} we show the best-fit prediction for the CMB multipoles. It is interesting to observe that, at low multipoles, the $R\Box^{-2}R$ model has a smaller amplitude, which goes in the direction indicated by the data, although of course in this region cosmic variance dominates. A similar result has been shown in  \cite{Barreira:2014kra} (although without performing parameter estimation for the model).

The conclusion of this analysis is that these non-local models, and particularly the $g_{\mu\nu}\Box^{-1}R$ model,
fit the present cosmological data  as well as $\Lambda$CDM, and in fact even slightly better if $H_0$ will turn out to have a high value. Both models have the same number of parameters as $\Lambda$CDM, and in this sense they are quite unique, among the vast literature on modified gravity models. To the best of our knowledge, there is no other model that is competitive with $\Lambda$CDM from the point of view of fitting current observations (at a level of accuracy which tests not only the background evolution but also the cosmological perturbations of the model), without being  an extension of $\Lambda$CDM with extra free parameters. This non-local approach therefore seems to provide an interesting new line of attack to the problem of finding a dynamical explanation for dark energy. From the point of view of the analysis of cosmological data, these non-local models also provide a welcome competitors, against which we can test the validity of $\Lambda$CDM.

\clearpage

\vspace{5mm}
\noindent
{\bf Acknowledgments.}
We thank Benjamin Audren for providing us a preliminary  version of Montepython including JLA data,  Alex Barreira for comparison of our codes for the $C_l$ for the $R\Box^{-2}R$ model, and Julien Lesgourgues for useful discussions.
The work of YD, SF, MK and MM is supported by the Fonds National Suisse. VP acknowledges the DFG TransRegio
TRR33 grant on ÔThe Dark UniverseÕ.

The development of Planck has been supported by: ESA; CNES and CNRS/INSU-IN2P3-INP (France); ASI, CNR, and INAF (Italy); NASA and DoE (USA); STFC and UKSA (UK); CSIC, MICINN and JA (Spain); Tekes, AoF and CSC (Finland); DLR and MPG (Germany); CSA (Canada); DTU Space (Denmark); SER/SSO (Switzerland); RCN (Norway); SFI (Ireland); FCT/MCTES (Portugal).
A description of the Planck Collaboration and a list of its members, including the technical or scientific activities in which they have been involved, can be found at 
http://www.cosmos.esa.int/web/planck/planck-collaboration.

\bibliographystyle{utphys}
\bibliography{myrefs_massive}

\end{document}